\documentclass[prl,preprint,amsmath,amssymb]{revtex4}
\usepackage{graphicx}
\usepackage[colorlinks,
            linkcolor=blue,
            anchorcolor=blue,
            citecolor=blue]{hyperref}
\newtheorem{theorem}{Theorem}

\def\ra{\rangle}
\def\la{\langle}
\allowdisplaybreaks[4]

\begin{document}

\title{A sufficient Entanglement Criterion Based On Quantum Fisher Information and Variance}

\author{Qing-Hua Zhang$^{1}$}
\author{Shao-Ming Fei$^{1,2}$}

\affiliation{$^1$School of Mathematical Sciences, Capital Normal University,
Beijing 100048, China\\
$^2$Max-Planck-Institute for Mathematics in the Sciences, 04103 Leipzig, Germany}

\bigskip

\begin{abstract}
We derive criterion in the form of inequality based on quantum Fisher information and quantum variance to detect multipartite entanglement. It can be regarded as complementary of the well-established PPT criterion in the sense that it can also detect bound entangled states. The inequality is motivated by Y.Akbari-Kourbolagh $et\ al.$[Phys. Rev A. 99, 012304 (2019)] which introduced a multipartite entanglement criterion based on quantum Fisher information. Our criterion is experimentally measurable for detecting any $N$-qudit pure state mixed with white noisy. We take several examples to illustrate that our criterion has good performance for detecting certain entangled states.
\end{abstract}

\maketitle

\section{\expandafter{\romannumeral1}. INTRODUCTION}
Entanglement, as a mysterious phenomenon of quantum mechanics, was first expounded by Einstein, Podolsky, and Rosen \cite{abn} and Schrödinger \cite{ed} questioning the completeness of quantum mechanics theory. Later, Bell noticed that entanglement caused experimentally testable deviations of quantum mechanics versus classical physics \cite{js}. Finally, with the development of quantum resource theory, entanglement was also recognized as a resource, which made sense of many quantum-information processing like quantum teleportation \cite{cgc}, quantum cryptography \cite{ak}, or measurement based quantum computation \cite{rh}. A natural problem is how to detect and quantify quantum entanglement. Unfortunately, it is still an open problem of determining whether any given quantum state is entangled or separable. This difficult problem has motivated the development of quantum information theory in the last decades \cite{rpmk}.

Up to now, many kinds of entanglement criteria have been proposed for the detection of entanglement in arbitrary quantum systems. But developing simpler and more efficient criterion is indeed pivotal. The most remarkable one for detection of entanglement is positive partial transpose (PPT) criterion proposed by Peres-Horodecki \cite{pa}. The condition is sufficient and necessary to characterize entanglement for $2\times 2$ and $2\times 3$ system \cite{mpr}, but in higher dimensional systems some entangled states escape the detection. Then Much subsequent work has been devoted to search more sufficient conditions for the entanglement detection of multipartite or higher dimension systems. The most remarkable one is the computable cross norm \cite{or} or realignment \cite{kl} (CCNR) criterion which have more powerful PPT entanglement detection capability. Another different type criteria are based on quantum Fisher information, located at the centre of quantum metrology \cite{pnlg,gt,pfd}, which are also complement the approaches based on variance and uncertainty relations \cite{hs}.

Recently, Y.Akbari-Kourbolagh $et$ $al.$ \cite{ym} proposed entanglement criterion for multipartite quantum systems based on quantum fisher information which had better performance than some other criteria. Considering from the aspect of detecting bound entangled states, it also complements the well-established PPT criterion. Moreover, the well-known upper bound of quantum Fisher information for separable multi-qudit states introduced by Toth and Vertesi \cite{gt} can be also improved by their method. But this method can not be experimentally measured. To overcome this flaw, we derive entanglement criterion in the form of inequality. Violation of this inequality reveals that the state is entangled and accordingly shows the usefulness for quantum metrology. We take several examples to illustrate that our criterion has good performance for detecting certain entangled states.

\section{\expandafter{\romannumeral2}. ENTANGLEMENT CRITERION}
The variance of a quantum mechanical observable $\hat{A}$ with respect to a quantum state  $\rho$ is defined by
\begin{equation}
(\bigtriangleup \hat{A})^2_{\rho}=\la \hat{A}^2\ra_{\rho} -\la \hat{A}\ra ^2 _{\rho}.
\end{equation}
The corresponding quantum Fisher information $F(\rho, \hat{A})$ is defined by \cite{ak,sl}
\begin{equation}\label{prefisher1}
F(\rho, \hat{A})=\frac{1}{4}Tr(\rho \hat{L}^2),
\end{equation}
where $\hat{L}$, the symmetric logarithmic derivative, is a Hermitian operator satisfying
$$
i[\rho,A]=\frac{1}{2}(\hat{L}\rho+\rho \hat{L}).
$$
Under the spectral decomposition of $\rho$, $\rho=\sum_k \lambda_k|k\ra \la k|$,
where $\lambda_k$ are eigenvalues and $|k\ra$ are the corresponding eigenvectors of $\rho$, the quantum Fisher information (\ref{prefisher1}) can be rewritten as:
\begin{equation}\label{qfi1}
F(\rho, \hat{A})=\sum_{k,l}\frac{(\lambda_k-\lambda_l)^2}{2(\lambda_k+\lambda_l)}|\la k|\hat{A}|l\ra|^2,
\end{equation}
where $\lambda_k+\lambda_l$ are nonzero \cite{pnlg,ak,sl,slcm,gt,pfd}.

For pure states, the quantum Fisher information equals to variance,
$F(\rho,\hat{A})=(\bigtriangleup \hat{A})_{\rho}^2$. For mixed state, the quantum Fisher information can be given by the convex roof of the variance \cite{gi,gd,sy,ym},
$$
F(\rho,\hat{A})=\inf_{\{p_k,|\psi_k\ra\}}\sum_kp_k(\bigtriangleup \hat{A})_{\psi_k}^2.
$$
where the infimum is taken over all possible pure state decompositions of $\rho=\sum_kp_k|\psi\ra\la\psi_k|$.
The quantum variance of mixed states can also be given by the convex roof of variances:
$$
(\bigtriangleup \hat{A})^2_{\rho}=\sup_{\{p_k,|\psi_k\ra\}}\sum_kp_k(\bigtriangleup \hat{A})_{|\psi_k\ra}^2.
$$
where the supremum is taken over all possible pure state decompositions of $\rho=\sum_kp_k|\psi\ra\la\psi_k|$.
Obviously, quantum information and variance satisfy the following inequality,
\begin{equation}\label{preeq1}
F(\rho,\hat{A})\leq\sum_kp_k(\bigtriangleup \hat{A})_{|\psi_k\ra}^2\leq(\bigtriangleup \hat{A})^2_{\rho}.
\end{equation}

Let $H_X$ be the Hilbert space associated to quantum system $X$ and $I$ be the identity matrix of $H_X$. Any bipartite separable state $\rho\in H_A\otimes H_B$ can be written as:
\begin{equation}\label{ecrho}
\rho=\sum_k{p_k|e_k\ra\la e_k|\otimes |f_k\ra\la f_k|},
\end{equation}
where $0<p_k\leq 1$, $\sum_k p_k=1$, $|e_k\ra$ and $|f_k\ra$ are arbitrary but normalized states of
the subsystems.

\begin{theorem}
Any separable bipartite state (\ref{ecrho}) must satisfy the following inequality:
\begin{equation}\label{th1}
F(\rho,\hat{A}\otimes I+I\otimes \hat{B})\leq \bigtriangleup (\hat{A}\otimes I-I\otimes \hat{B})^2_{\rho}
\end{equation}
for any local observables $\hat{A}$ and $\hat{B}$.
\end{theorem}

{\sf [Proof]} For any separable state (\ref{ecrho}), by the inequality (\ref{preeq1}), we get
\begin{equation}\label{th1eq1}
\begin{aligned}
F(\rho,\hat{A}\otimes I+I\otimes \hat{B})\leq &\sum_{k}p_k\bigtriangleup(\hat{A}+\hat{B})_{|e_k\ra\otimes|f_k\ra}^2   \\
=&\la \hat{A}^2\ra_{\rho_A}+\la \hat{B}^2\ra_{\rho_B}-\sum_kp_k[\la e_k|\hat{A}|e_k\ra^2+\la f_k|\hat{B}|f_k\ra^2] \\
=&\la \hat{A}^2\ra_{\rho_A}+\la \hat{B}^2\ra_{\rho_B}-2\sum_kp_k \la\hat{A}\ra_{|e_k \ra}\la\hat{B}\ra_{|f_k \ra}  \\
&-\sum_kp_k[\la e_k|\hat{A}|e_k\ra-\la f_k|\hat{B}|f_k\ra]^2                                                       \\
\leq &\la(\hat{A}\otimes I-I\otimes \hat{B})^2\ra_{\rho}-[\sum_kp_k\la (e_k|\hat{A}|e_k\ra-\la f_k|\hat{B}|f_k\ra)]^2                 \\
=&\la(\hat{A}\otimes I-I\otimes \hat{B})^2\ra_{\rho}-\la(\hat{A}\otimes I-I\otimes \hat{B})\ra_{\rho}^2                            \\
=&\bigtriangleup (\hat{A}\otimes I-I\otimes \hat{B})^2_{\rho}.
\end{aligned}
\end{equation}
\hfill \rule{1ex}{1ex}

From the inequality (\ref{preeq1}), we can obtain that the inequality $F(\rho,\hat{A}\otimes I+I\otimes \hat{B})\leq \bigtriangleup (\hat{A}\otimes I+I\otimes \hat{B})^2_{\rho}$ holds for all quantum states. From Theorem 1 we have that (\ref{th1}) holds for all separable states. If for some proper observables such that the state $\rho$ satisfies $Cov_\rho(\hat{A}\otimes I,I \otimes \hat{B})=\la \hat{A}\otimes \hat{B}\ra -\la \hat{A}\otimes I\ra\la I\otimes \hat{B}\ra>0$, then from (\ref{th1}) the state $\rho$ is entangled.

In \cite{ym} an elegant entanglement criterion has been derived: For any bipartite separable state (\ref{ecrho}), the following inequality must hold,
\begin{equation}\label{ym1}
F(\rho,\hat{A}\otimes I+I\otimes \hat{B})\leq \la\hat{A}\ra^2_{\rho_A}+\la\hat{B}^2\ra_{\rho_B}-2|\langle\hat{A}\otimes\hat{B}\rangle_{\rho}|,
\end{equation}
where $\rho_A$ and $\rho_B$ are the reduced states of $\rho$ associated with the subsystems $A$ and $B$, respectively.

\emph{Example 1.} Consider the two-ququart PPT bound entangled state \cite{ym},
\begin{equation*}
\rho=p\sum_{n=1}^{4}|\psi_n\ra\la\psi_n|+q\sum_{n=5}^{6}|\psi_n\ra\la\psi_n|,\qquad 4p+2q=1,
\end{equation*}
where p and q are non-negative real numbers, the pure states $|\psi_n\ra$ are defined by,
\begin{equation*}
\begin{aligned}
&|\psi_1\ra=\frac{1}{2}(|01\ra+|23\ra),\qquad  |\psi_2\ra=\frac{1}{2}(|10\ra+|32\ra),\\
&|\psi_3\ra=\frac{1}{2}(|11\ra+|22\ra),\qquad  |\psi_4\ra=\frac{1}{2}(|00\ra-|33\ra),\\
&|\psi_5\ra=\frac{1}{2}(|03\ra+|12\ra)+\frac{1}{\sqrt{2}}|21\ra,\\
&|\psi_6\ra=\frac{1}{2}(-|03\ra+|12\ra)+\frac{1}{\sqrt{2}}|30\ra.
\end{aligned}
\end{equation*}
Set $p=(2-\sqrt{2})/4$ and $q=(\sqrt{2}-1)/2$. The state is then a metrologically useful PPT bound entangled state \cite{hs}.
Taking observables $\hat{A}=\hat{B}= |0\ra\la0|+|1\ra\la1|-|2\ra\la2|-|3\ra\la3|$, we have
\begin{equation*}
F(\rho,\hat{A}\otimes I+I\otimes \hat{B})=8-4\sqrt{2},
\end{equation*}
and
\begin{equation*}
\bigtriangleup (\hat{A}-\hat{B})^2_{\rho}=4\sqrt{2}-4.
\end{equation*}
Obviously inequality (\ref{th1}) is violated. Therefore, from Theorem 1 the entanglement is detected, which is in accordance with the result from \cite{ym}.

The state in above example is isotropic and the observables for each system have been taken to be same ones. Now let us consider an non-isotropic state.

\emph{Example 2.} Consider a two-qubit state with white noise:
\begin{equation*}
\rho=p|\psi\ra\la\psi|+(1-p)\frac{I\otimes I}{4},\qquad 0\leq p\leq 1,
\end{equation*}
where $|\psi\ra=\frac{2}{3}(|00\ra+|11\ra)+\frac{1}{3}|10\ra$. From the PPT criterion, this state is entangled if and only if $p>\frac{9}{25}$.

From the fact that for a $N$-qudit quantum state $|\psi\ra$ mixed with white noise,
$\rho_{noisy}(p)=p|\psi\ra\la\psi|+(1-p)\frac{I^{\otimes N}}{d^N}$, the
quantum Fisher information of $\rho_{noisy}(p)$ can be rewritten as \cite{pnlg}
$$
F(\rho_{noisy},\hat{A})=\frac{p^2}{p+2(1-p)d^{-N}} F(|\psi\ra,\hat{A}),
$$
For this state we obtain
\begin{equation*}
F(\rho,\hat{A}\otimes I+I\otimes \hat{B})=\frac{64p^2}{9p + 9},
\end{equation*}
where we have taken $\hat{A}=\hat{B}=\sigma_z$.
It is also direct to verify that
\begin{equation*}
\bigtriangleup (\hat{A}-\hat{B})^2_{\rho}=2 - \frac{4p^2}{81} - \frac{14p}{9}.
\end{equation*}
Hence, $\rho$ violates the inequality (\ref{th2}) for $p>0.5044$.

From (\ref{ym1}) given in \cite{ym}, one can get
\begin{equation*}
\la\hat{A}^2\ra_{\rho_{A}}-\la\hat{B}^2\ra_{\rho_{B}}-|\la\hat{A}\otimes \hat{B}\ra_{\rho}|=2 - \frac{14p}{9},
\end{equation*}
namely, $\rho$ violates the inequality for $p>0.5067$. Therefore, our Theorem 1 detects better entanglement than (\ref{ym1}) in this case.

Our results can be generalized to tripartite systems. Any fully separable tripartite state $\rho\in H_A\otimes H_B\otimes H_C$ has the following form,
\begin{equation}\label{rho3}
 \rho=\sum_k{p_k|e_k\ra\la e_k|\otimes |f_k\ra\la f_k|\otimes|g_k\ra\la g_k|}.
\end{equation}
where $0<p_k\leq 1$, $\sum_k p_k=1$, $|e_k\ra$, $|f_k\ra$ and $|g_k\ra$ are normalized states of the three subsystems.

\begin{theorem}
Any separable state (\ref{rho3}) must satisfy the following inequality,
\begin{equation}\label{th2}
F(\rho,\hat{A}+\hat{B}+\hat{C})\leq \frac{1}{2}[\bigtriangleup (\hat{A}-\hat{B})^2_{\rho_{AB}}+\bigtriangleup (\hat{A}-\hat{C})^2_{\rho_{AC}}+\bigtriangleup (\hat{B}-\hat{C})^2_{\rho_{BC}}].
\end{equation}
where $\hat{A}+\hat{B}+\hat{C}$ stands for $\hat{A}\otimes I_{BC}+\hat{B}\otimes I_{AC}+ \hat{C}\otimes I_{AB}$, $\hat{A}-\hat{B}$ denotes $\hat{A}\otimes I-I\otimes\hat{B}$ and $\rho_{AB}=tr_C(\rho)$,
$\hat{A}-\hat{C}$, $\hat{B}-\hat{C}$, $\rho_{AC}$ and $\rho_{BC}$ are similarly defined.
\end{theorem}

{\sf [Proof]} For any separable state (\ref{rho3}), from the inequality (\ref{preeq1}), we have
\begin{align*}
F(\rho,\hat{A}+\hat{B}+\hat{C})\leq &\sum_{k}p_k\bigtriangleup(\hat{A}+\hat{B}+\hat{C})_{|e_k\ra\otimes|f_k\ra\otimes|g_k\ra}^2   \\
=&\sum_{k}p_k[(\bigtriangleup\hat{A})_{|e_k\ra}^2+(\bigtriangleup \hat{B})_{|f_k\ra}^2+(\bigtriangleup \hat{C})_{|g_k\ra}^2]\\
=&\la \hat{A}^2\ra_{\rho_A}\!+\!\la\hat{B}^2\ra_{\rho_B}\!+\!\la\hat{C}^2\ra_{\rho_C}-\sum_kp_k[\la\hat{A}\ra_{|e_k\ra}^2+\la\hat{B}\ra_{|f_k\ra}^2+\la\hat{C}\ra_{|g_k\ra}^2] \\
=&\frac{1}{2}\la \hat{A}^2\ra_{\rho_A}\!+\!\frac{1}{2}\la \hat{B}^2\ra_{\rho_B}\!-\!\la\hat{A}\otimes\hat{B}\ra_{\rho_{AB}}\!-\!\frac{1}{2}\sum_kp_k[\la\hat{A}\ra_{|e_k\ra}\!-\!\la\hat{B}\ra_{|f_k\ra}]^2 \\
&\!+\! \frac{1}{2}\la \hat{A}^2\ra_{\rho_A}\!+\!\frac{1}{2}\la \hat{C}^2\ra_{\rho_C}\!-\!\la\hat{A}\otimes\hat{C}\ra_{\rho_{AC}}\!-\!\frac{1}{2}\sum_kp_k[\la\hat{A}\ra_{|e_k\ra}\!-\!\la\hat{C}\ra_{|g_k\ra}]^2 \\
&\!+\! \frac{1}{2}\la \hat{B}^2\ra_{\rho_B}\!+\!\frac{1}{2}\la \hat{C}^2\ra_{\rho_C}\!-\!\la\hat{B}\otimes\hat{C}\ra_{\rho_{BC}}\!-\!\frac{1}{2}\sum_kp_k[\la\hat{B}\ra_{|f_k\ra}\!-\!\la\hat{C}\ra_{|g_k\ra}]^2 \\
\leq &\frac{1}{2}\la(\hat{A}\otimes I-I\otimes \hat{B})^2\ra_{\rho_{AB}}-\frac{1}{2}[\sum_kp_k(\la\hat{A}\ra_{|e_k\ra}-\la\hat{B}\ra_{|f_k\ra)})]^2\\
&+\frac{1}{2}\la(\hat{A}\otimes I-I\otimes\hat{C})^2\ra_{\rho_{AC}}-\frac{1}{2}[\sum_kp_k(\la\hat{A}\ra_{|e_k\ra}-\la\hat{C}\ra_{|g_k\ra})]^2\\
&+\frac{1}{2}\la(\hat{B}\otimes I-I\otimes \hat{C})^2\ra_{\rho_{BC}}-\frac{1}{2}[\sum_kp_k(\la\hat{B}\ra_{|f_k\ra}-\la\hat{C}\ra_{|g_k\ra})]^2  \\
=&\frac{1}{2}[\bigtriangleup (\hat{A}-\hat{B})^2_{\rho_{AB}}+\bigtriangleup (\hat{A}-\hat{C})^2_{\rho_{AC}}+\bigtriangleup (\hat{B}-\hat{C})^2_{\rho_{BC}}].
\end{align*}
\hfill \rule{1ex}{1ex}

\begin{figure}
  \centering
  \includegraphics[width=11cm]{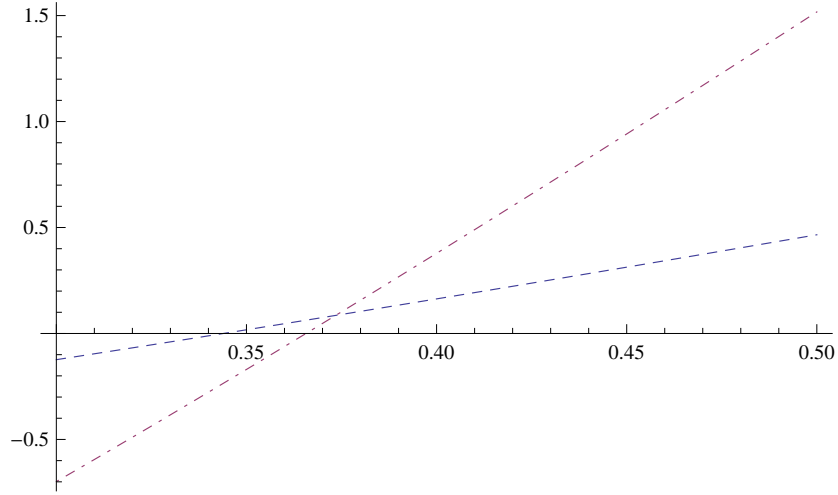}
  \caption{Red (dashed) line represents our criterion for Example 3, and blue (dot-dashed) line represent violation of inequality (\ref{ym2}) given in ref.\cite{ym}}
  \label{fig1}
\end{figure}

\emph{Example 3.} Consider a three-qubit GHZ-type state mixed with white noise \cite{ysw},
\begin{equation}\label{ex3}
\rho=p|\psi\ra\la\psi|+(1-p)\frac{I\otimes I \otimes I}{8},\qquad 0\leq p\leq 1,
\end{equation}
where $|\psi\ra=\frac{2}{3}(|000\ra+|111\ra)+\frac{1}{3}|110\ra$. Taking $\hat{A}=\hat{B}=-|1\ra\la1|$ and $\hat{C}=|0\ra\la0|$, we have
\begin{equation*}
F(\rho,\hat{A}+\hat{B}+\hat{C})=\frac{656p^2}{243p + 81},
\end{equation*}
and
\begin{equation*}
\frac{1}{2}[\bigtriangleup (\hat{A}-\hat{B})^2_{\rho_{AB}}+\bigtriangleup (\hat{A}-\hat{C})^2_{\rho_{AC}}+\bigtriangleup (\hat{B}-\hat{C})^2_{\rho_{BC}}]=\frac{7}{4}-(\frac{p}{9}+1)^2 - \frac{7p}{12}
\end{equation*}
the violation of the inequality (\ref{th2}) is equivalent to solve the following equality:
\begin{equation*}
y1:=\frac{656p^2}{243p + 81}-(\frac{7}{4}-(\frac{p}{9}+1)^2 - \frac{7p}{12})>0
\end{equation*}
Hence $\rho$ violates the inequality (\ref{th2}) for $p>0.3439$.
An elegant entanglement criterion for tripartite system has been derived \cite{ym}: For any tripartite separable state (\ref{rho3}), the inequality must satisfied:
\begin{equation}\label{ym2}
F(\rho,\hat{A}+\hat{B}+\hat{C})\leq \la \hat{A}^2\ra_{\rho_A}+\la \hat{B}^2\ra_{\rho_B}+\la \hat{C}^2\ra_{\rho_C}-\eta,
\end{equation}
where $\eta=|\la\hat{A}\otimes\hat{B}\ra_{\rho_{AB}}|+|\la\hat{B}\otimes\hat{C}\ra_{\rho_{BC}}|+|\la\hat{A}\otimes\hat{C}\ra_{\rho_{AC}}|$. Taking the observables $\hat{A}=\hat{B}=\hat{C}=\sigma_z$, we have:
\begin{equation*}
F(\rho,\hat{A}+\hat{B}+\hat{C})=\frac{2624p^2}{243p + 81},
\end{equation*}
the right hand of inequality (\ref{ym2}) can be also derived:
\begin{equation*}
\la\hat{A}^2\ra_{\rho_A}+\la \hat{B}^2\ra_{\rho_B}+\la \hat{C}^2\ra_{\rho_C}-\eta=3 - \frac{23p}{9},
\end{equation*}
the violation of the inequality (\ref{ym2}) is equivalent to solve the following equality:
\begin{equation*}
y2:=\frac{2624p^2}{243p + 81}-(3 - \frac{23p}{9})>0,
\end{equation*}
By simple calculation, $\rho$ violates the inequality (\ref{ym2}) for $p>0.3657$.
Both results are better than the one, $p>\frac{9}{23}$, based on inequality relation of the mean values of certain observables \cite{y}.
Therefore, our criterion has better performance than the one presented in \cite{ym} in detecting the entanglement of the non-isotropic state (\ref{ex3}), see Fig. \ref{fig1}.

Our conclusions can be further generalized to multipartite quantum systems.
For arbitrary multipartite quantum system $A_1A_2\ldots A_N$, any fully separable state $\rho\in H_A\otimes H_B\otimes\ldots\otimes H_N$ can be represented as
\begin{equation}\label{rhoN}
\rho=\sum_k{p_k|e_k\ra^{1}\la e_k|\otimes |e_k\ra^{2}\la e_k|\otimes\ldots\otimes|e_k\ra^{N}\la e_k|}.
\end{equation}
where $0<p_k\leq 1$, $\sum_k p_k=1$, $|e_k\ra^i,i=1,2,\ldots,N$ are normalized states of $i$ subsystems.
Similar to the form of Theorem 2, we have the following theorem for multipartite quantum systems.

\begin{theorem}               
Any separable state (\ref{rhoN}) must satisfy the inequality,
\begin{equation}\label{th3}
F(\rho,\hat{A_1}+\hat{A_2}+\cdots +\hat{A_N})\leq \frac{1}{2}\sum_{i=1}^N\bigtriangleup (\hat{A_i}-\hat{A}_{i+1})^2_{\rho_{A_iA_{i+1}}},
\end{equation}
where $\hat{A}_{N+1}=\hat{A_1}$. violation of the inequality implies $\rho$ is entangled.
\end{theorem}
The proof of this theorem is similar to theorem 2, so we omit here.

\section{\expandafter{\romannumeral3}. CONCLUSION}
We proposed a sufficient entanglement criterion for the multipartite systems based on quantum Fisher information and variance of local observables. Our criterion have better performance for detecting non-isotropic states. And our criterion is experimentally measurable for detecting any $N$-qudit pure state mixed with white noisy. We calculated several examples to illustrate the efficiency. In \citep{dp,pfd,dc}, it has been shown that it is possible to establish connection between generalized variances and generalized quantum Fisher information. For more generalized definition of quantum Fisher information, maybe one can derive more efficient results. It would be also interesting to deal with the problem of detecting the genuine multipartite entanglement based on quantum Fisher information, which will be our next considerations. Our results may highlight further investigations on both the separability of bipartite states and the genuine multipartite entanglement.

\bigskip
\noindent{\bf Acknowledgments}\, \,  This work is supported by the NSF of China under Grant No. 11675113, and Beijing Municipal Commission of Education (KZ201810028042).


\begin{thebibliography}{99}
\bibitem[1]{abn} A. Einstein, B. Podolsky, N. Rosen, Can quantum-mechanical description of physical reality be considered complete? Phys. Rev. 47, 777 (1935).
\bibitem[2]{ed} E. Schrödinger, Die gegenwärtige Situation in der Quantenmechanik, Die Naturwissenschaften 23 (1935) 807–812; 823–828; 844–849.
\bibitem[3]{js} J.S. Bell, On the einstein podolsky rosen paradox, Physics 1, 195 (1964). Reprinted in J. Bell, Speakable and Unspeakable in Quantum Mechanics, Cambridge University Press (2004).
\bibitem[4]{cgc} C.H. Bennett, G. Brassard, C. Crépeau, R. Jozsa, A. Peres, W.K. Wootters, Teleporting an unknown quantum state via dual classical and Einstein-Podolsky-Rosen channels, Phys. Rev. Lett. 70, 1895 (1993).
\bibitem[5]{ak} A.K. Ekert ,Quantum cryptography based on Bell's theorem, Phys. Rev. Lett. 67, 661 (1991).
\bibitem[6]{rh} R. Raussendorf, H.J. Briegel, A one-way quantum computer, Phys Rev. Lett. 86, 5188 (2001).
\bibitem[7]{rpmk} R. Horodecki, P. Horodecki, M. Horodecki, and K. Horodecki, Quantum entanglement, Rev. Mod. Phys. 81, 865 (2009); O. Gühne and G. Tóth, Entanglement detection, Phys. Rep. 474, 1 (2009).
\bibitem[8]{pa} A. Peres, Separability criterion for density matrices, Phys. Rev. Lett. 77, 1413 (1996).
\bibitem[9]{mpr} M. Horodecki, P. Horodecki and R. Horodecki, Teleportation, Bell's inequalities and inseparability, Phys. Lett. A 223, 1 (1996).
\bibitem[10]{or} O. Rudolph, Further results on the cross norm criterion for separability, Quantum Inf Process 4: 219 (2005).
\bibitem[11]{kl} K. Chen and L.A. Wu, A matrix realignment method for recognizing entanglement, Quantum Inf. Comput. 3, 193 (2003).
\bibitem[12]{pnlg}P. Hyllus, W. Laskowski, R. Krischek, C. Schwemmer, W. Wieczorek, H. Weinfurter, L. Pezze, and A. Smerzi, Fisher information and multiparticle entanglement, Phys. Rev. A 85, 022321 (2012); N. Li and S. Luo, Entanglement detection via quantum Fisher information, Phys. Rev. A 88, 014301 (2013).
\bibitem[13]{gt}G. Toth and T. Vertesi, Quantum states with a positive partial transpose are useful for metrology, Phys. Rev. Lett. 120, 020506 (2018)
\bibitem[14]{pfd} P. Gibilisco, F. Hiai, and D. Petz, Quantum covariance, quantum Fisher information, and the uncertainty relations, IEEE Trans. Inf. Theory 55, 439 (2009).
\bibitem[15]{hs} H. F. Hofmann and S. Takeuchi, Violation of local uncertainty relations as a signature of entanglement, Phys. Rev. A 68, 032103 (2003).
\bibitem[16]{ym} Y. Akbari-Kourbolagh and M. Azhdargalam, Entanglement criterion for multipartite systems based on quantum Fisher information, Phys. Rev A. 99, 012304 (2019).
\bibitem[17]{gi} G. Toth and I. Apellaniz, Quantum metrology from a quantum information science perspective, J. Phys. A 47, 424006 (2014).
\bibitem[18]{gd} G. Toth and D. Petz, Extremal properties of the variance and the quantum Fisher information, Phys. Rev. A 87, 032324 (2013).
\bibitem[19]{sy} S. Yu, Quantum Fisher information as the convex roof of variance, arXiv:1302.5311.
\bibitem[10]{sl} S. Luo, Wigner-Yanase skew information vs. quantum Fisher information, Proc. Am. Math. Soc. 132, 885 (2004).
\bibitem[21]{slcm} S. L. Braunstein and C. M. Caves, Statistical distance and the geometry of quantum states, Phys. Rev. Lett. 72, 3439 (1994).
\bibitem[22]{y}Y. Akbari-Kourbolagh, Entanglement criteria for the three-qubit states, Int. J. Quantum Inf. 15, 1750049 (2017).
\bibitem[23]{ysw} Y. S. Weinstein, Tripartite entanglement witnesses and entanglement sudden death, Phys. Rev. A 79 012318 (2009).
\bibitem[24]{dp} D. Petz, Covariance and Fisher information in quantum mechanics, J. Phys. A 35, 929 (2002).
\bibitem[25]{dc} For a review on generalized variances, types of Fisher information, and covariances, see D. Petz and C. Ghinea, in Introduction to Quantum Fisher Information, QP-PQ: Quantum Probability and White Noise Analysis. Vol. 27, edited by R. Rebolledo and M. Orszag (World Scientific,Singapore, pp. 261–281 (2011).

\end{thebibliography}
\end{document}